\documentclass[letterpaper]{article} 
\usepackage{aaai2027}  
\usepackage[hyphens]{url}  
\usepackage{graphicx} 
\usepackage{natbib}  
\usepackage{caption} 
\usepackage{algorithm}
\usepackage{algorithmic}
\usepackage{tikz}
\usepackage{pgf}
\usepackage{pgfplots}
\usepackage{pgfplotstable}
\usetikzlibrary{positioning,arrows.meta,shapes.misc}

\usepackage{newfloat}
\usepackage{listings}
\DeclareCaptionStyle{ruled}{labelfont=normalfont,labelsep=colon,strut=off} 
\floatstyle{ruled}
\newfloat{listing}{tb}{lst}{}
\floatname{listing}{Listing}

\usepackage{booktabs}
\usepackage{amsmath,amssymb} 

\title{\textsc{PopMelodyGen}: A Copyright-Aware Platform for Controllable Pop-Melody Generation}
\author{
    Di Lu\textsuperscript{\rm 1,2,3}\equalcontrib,
    Xinran Zhang\textsuperscript{\rm 1,2,3}\equalcontrib,
    Qingwen Zhou\textsuperscript{\rm 1,2,3}\equalcontrib,
    Yan Gao\textsuperscript{\rm 1,2,3},
    Yuan Zhang\textsuperscript{\rm 1,2,3},
    Xiaoqing Wang\textsuperscript{\rm 1,2,3},
    Yuming Sun\textsuperscript{\rm 1,2,3},
    Xiaobing Li\textsuperscript{\rm 1,2,3},
    Maosong Sun\textsuperscript{\rm 1,2,3,4}\corresponding
}
\affiliations{
    \textsuperscript{\rm 1}Department of AI Music and Music Information Technology, Central Conservatory of Music\\
    \textsuperscript{\rm 2}Key Laboratory of Music and Brain Science, Central Conservatory of Music, Ministry of Education, China\\
    \textsuperscript{\rm 3}Laboratory of AI Music, Central Conservatory of Music, Laboratory of Philosophy and Social Sciences, Ministry of Education, China\\
    \textsuperscript{\rm 4}Department of Computer Science and Technology, Tsinghua University\\
    \{ld, zxinran, zqingwen\}@ccom.edu.cn, sms@tsinghua.edu.cn
}

\begin{document}

\maketitle

\begin{abstract}
Large-scale redistributable symbolic pop-melody data remain scarce due to copyright restrictions, creating a bottleneck for both practical deployment and reproducible research in music generation. 
We present \textsc{PopMelodyGen}, a copyright-aware platform for controllable pop-melody generation that combines rule-based synthesis with neural generation.
Our approach first uses an expert synthesizer to produce large-scale labeled synthetic melodies from explicit musical knowledge and templates, providing scalable pretraining data with auditable provenance.
A neural model is then pretrained on this corpus and adapted with a small licensed real-pop corpus to improve melodic naturalness while preserving controllability.
The platform supports interactive melody creation with accompaniment playback and export for creators, as well as batch construction of annotated symbolic datasets for researchers.
\textsc{PopMelodyGen} demonstrates a practical training and deployment paradigm for creative AI when large-scale reusable real-world data are limited.
\end{abstract}

 \begin{links}
     \link{Code}{https://github.com/ld04080106/PopMelodyGen}
     \link{Video}{https://youtu.be/U_R1MEZFOjI}
 \end{links}
\section{Introduction}
Recent advances in generative AI rely heavily on large-scale training corpora, yet this assumption breaks down for symbolic pop melodies, where commercial pop songs are subject to copyright restrictions and are difficult to redistribute at scale.

Existing systems explore controllable symbolic generation across different architectures \cite{Choi2021,huang2020remi,Tian2025XMusicTA,notegen2025}, but reproducible development still depends on suitable symbolic training corpora.
Available pop-oriented datasets such as POP909 and RWC \cite{Wang2020POP909AP,Goto2002RWC,RWC2026Zenodo} remain limited in scale and condition coverage for controllable pop-melody pretraining, while larger reusable symbolic corpora such as PDMX \cite{Phillip2025pdmx} are not specifically designed for this setting.

To address this bottleneck, we present \textsc{PopMelodyGen}, a copyright-aware platform built around synthetic pretraining and licensed real-pop adaptation.
An expert synthesizer first constructs labeled training data from general music-theoretic rules and musician-curated structure, harmony, and rhythm templates rather than existing pop melodies.
A neural model is pretrained on this synthetic corpus and subsequently adapted toward real-pop melodic patterns.
We deploy the resulting workflow as a web platform supporting both interactive melody creation and batch construction of annotated symbolic datasets, reducing copyright-related barriers to reproducible music-AI research.

\section{System Overview}
\begin{figure*}[t]
\centering
\includegraphics[width=1\textwidth]{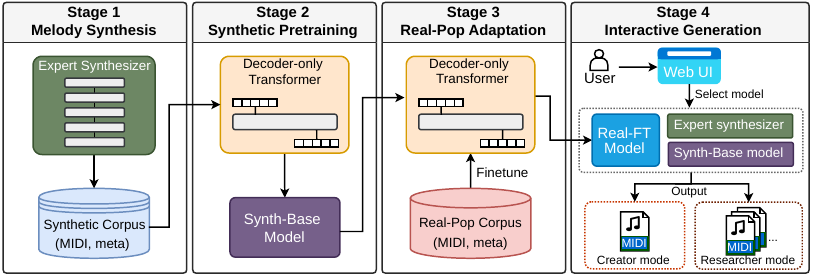} 
\caption{
Overview of the \textsc{PopMelodyGen} pipeline, from synthetic data generation to interactive deployment.
}
\label{fig}
\end{figure*}

Figure~\ref{fig} shows the four-stage pipeline of \textsc{PopMelodyGen}.

\paragraph{Synthetic melody generation.}
Our implemented expert synthesizer generates pop‑style melodies in a standard symbolic music format (MIDI) under explicit musical controls: tempo class, key/mode, vocal register, song structure, and chord progression.
Each output is paired with structured metadata capturing generation settings and annotations.
Based on general music‑theoretic rules and musician‑curated element‑level templates (structure, harmony, rhythm), 
the synthesizer does not ingest or transform existing pop melodies, giving each generated example fully inspectable provenance.

\paragraph{Synthetic pretraining.}
The generated MIDI--metadata pairs are converted into event-based token representations, following established symbolic music encoding practices \cite{huang2020remi}.
A decoder-only Transformer is pretrained with next-token prediction to learn controllable autoregressive melody generation, yielding the \textsc{Synth-Base} model.

\paragraph{Real-pop adaptation.}
Because rule-generated melodies may differ from human-composed pop music, we adapt \textsc{Synth-Base} with a small licensed real-pop corpus.
Real-pop sequences are mixed with sampled synthetic replay examples to preserve controllability, yielding the \textsc{Real-FT} model.

\paragraph{Interactive generation.}
The web platform passes the user-selected musical conditions to the chosen generation backend.
The resulting melody is represented in MIDI for visualization, playback, and export through the web platform.

\section{Live Demonstration}
The live demonstration exposes two complementary workflows. 
In \textbf{Creator Mode}, an attendee selects a generation backend (\textsc{Expert} / \textsc{Synth-Base} / \textsc{Real-FT}) and specifies musical conditions including key/mode, tempo class, vocal register, song structure, and chord progression. 
The system generates a pop-style melody together with built-in drums,
bass, piano, and guitar accompaniment for immediate audition.
Attendees can regenerate the same musical conditions with different backends, compare their outputs, and export the result for further editing.

In \textbf{Researcher Mode}, attendees can configure the range of musical conditions (e.g., 12 keys × 3 tempo classes) and batch-generate labeled MIDI--metadata pairs. 
The structured dataset package can be downloaded directly.
This demonstrates how the same system can serve not only as a creative tool, but also as a data-generation platform for reproducible music-AI experiments.

\section{Implementation and Validation}

\paragraph{Implementation.} 
For synthetic pretraining, we use 100k rule-generated melodies to pretrain the \textsc{Synth-Base} model.
For real-pop adaptation, we use the licensed RWC Popular Music Database \cite{Goto2002RWC,RWC2026Zenodo}.
After preprocessing and transposition-based augmentation, 98 prepared RWC melodies yield 847 tokenized adaptation examples.
Both stages use a 6-layer decoder-only Transformer with 512-dimensional hidden states and 8 attention heads.
The Web interface is implemented with Node.js and deployed on a dedicated server, with model inference served through a backend API.

\paragraph{Validation.}
For objective evaluation, each of \textsc{Expert}, \textsc{Synth-Base}, and \textsc{Real-FT} generates 4,320 melodies under systematically varied musical conditions.
\textsc{Synth-Base} approaches the expert synthesizer in harmony adherence and cadence control, whereas \textsc{Real-FT} shifts toward the RWC reference statistics while retaining much of the learned controllability.
In a human study with 30 participants, \textsc{Real-FT} achieves the highest MOS (3.55/5) and best-of-three preference rate (48.3\%), suggesting improved perceived melodic quality after real-pop adaptation.


\section{Responsible Use}

We describe \textsc{PopMelodyGen} as copyright-aware rather than legally copyright-safe.
The synthetic corpus is constructed from general musical rules and element-level templates without ingesting or transforming existing pop melodies, while real-pop data are introduced only through the licensed RWC dataset.

As an empirical sanity check, we compare 12,960 generated melodies against the RWC reference set using interval n-gram nearest-neighbor similarity, inspired by symbolic melodic-similarity methods using local fragments and transposition-invariant interval representations \cite{Urbano2014MelodyShape,Lattner2018IntervalFeatures}.
The mean nearest-neighbor similarity is 0.061 and the maximum is 0.336. 
Manual inspection of the highest-scoring pairs revealed no obvious near-duplicate melodies.
This analysis helps identify near-copy risks but does not constitute a legal guarantee.

\section{Conclusion}
\textsc{PopMelodyGen} addresses the training-data bottleneck in controllable pop-melody generation through pretraining on rule-generated synthetic data and adaptation with limited licensed real-pop data.
The resulting platform supports both interactive pop-melody creation and research-oriented dataset construction, illustrating a practical strategy for creative AI when large-scale reusable training data are unavailable.

\section*{Acknowledgments}
This research is funded by the Special Program of the National Natural Science Foundation of China (Grant No. T2341003) and the Advanced Discipline Construction Project of Beijing Universities.

\bibliography{popmelodygen}

\clearpage

\appendix
\setcounter{secnumdepth}{2} 

\section*{Appendix}

\providecommand{\PopMissingFigure}[1]{%
  \fbox{\begin{minipage}[c][4.0cm][c]{0.89\linewidth}
    \centering\small\textbf{Figure asset to be supplied}\par\vspace{6pt}
    {\ttfamily\footnotesize\detokenize{#1}}\par\vspace{6pt}
    \normalfont\footnotesize Place the original image at this path and recompile.
  \end{minipage}}%
}


\section{Expert Synthesizer}
\label{app:expert}

\begin{figure}[htbp]
\centering
\IfFileExists{Figures/fig2_expert.drawio_5.pdf}{%
  \includegraphics[width=\linewidth]{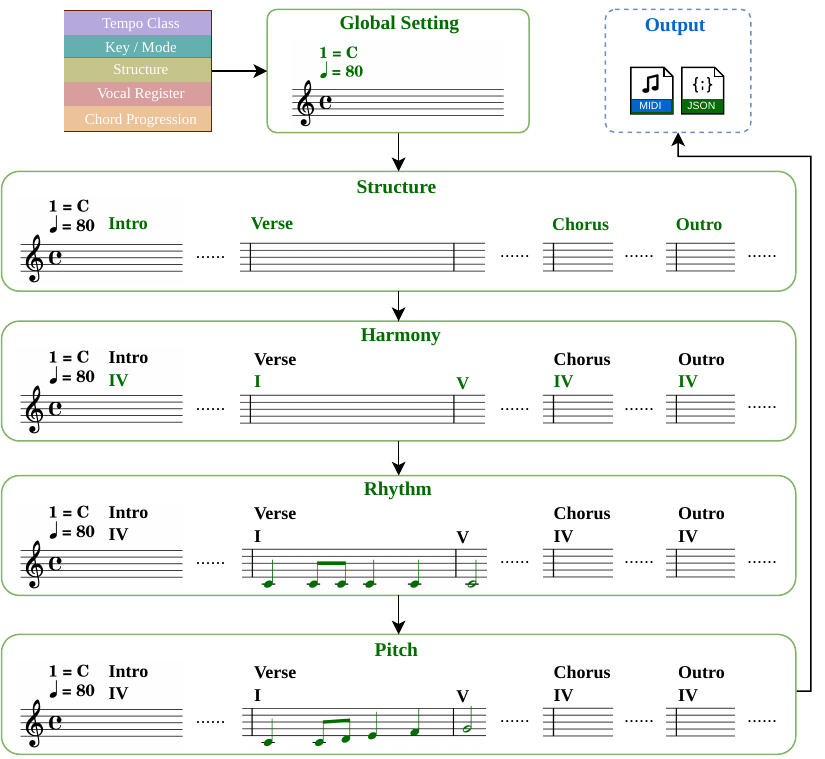}%
}{\PopMissingFigure{Figures/fig2_expert.drawio_5.pdf}}
\caption{Expert synthesizer for synthetic melody generation. The system instantiates global settings, structure, harmony, rhythm, and pitch, and outputs a melody MIDI file with structured metadata.}
\label{fig:fig2_pipeline}
\end{figure}

The purpose of the expert stage is to generate a large-scale, diverse, and controllable synthetic corpus for symbolic pop-melody pretraining. 
Figure~\ref{fig:fig2_pipeline} shows this sequence.

The synthesizer takes controllable musical conditions as input, including tempo class, key/mode, vocal register, structure, and chord progression. It first instantiates global settings, then a pop-song structure, followed by section-wise chord progressions expanded into bar-wise harmonic contexts. Rhythm is generated at the phrase level on an eighth-note grid, and pitches are sampled under harmony, register, and interval-transition constraints. Strong positions are biased toward chord tones, while weak positions are biased toward scale tones. Each generated example consists of a symbolic melody track and structured metadata describing tempo, key, register, section boundaries, and harmony.

\subsection{Global Conditions}
\label{app:global}

The global conditions include tempo class (Slow, Mid, or Fast), key, vocal register (Male or Female), song structure, and chord progression. The implementation used for synthetic pretraining enumerates 12 major keys, uses a fixed 4/4 meter, and samples additional musical conditions under the expert rules.

\subsection{Song Structure}
\label{app:structure}
The structure family is \emph{Intro--Verse--[Pre-chorus]--Chorus--Outro}. Intro and Outro are fixed to eight bars, and an optional Pre-chorus is fixed to eight bars. The Verse and Chorus candidates are shown in Table~\ref{tab:app_structures}. The Cartesian product of three Verse patterns, two Pre-chorus choices, and three Chorus patterns yields 18 structural templates. The implementation also filters candidate structures by a configurable song-duration range, set to 60--180 seconds in the reported implementation.

\begin{table}[htbp]
\centering
\small
\begin{tabular}{|c|c|c|c|c|}
\hline
\textbf{Intro} & \textbf{Verse} & \textbf{Pre-chorus} & \textbf{Chorus} & \textbf{Outro} \\
\hline
I8 & A8A8    & -  & B8B8    & O8 \\
   & A16     & C8 & B16     &  \\
   & A16A16  &    & B16B16  &  \\ 
\hline
\end{tabular}
\caption{Structure-template families. I, A, C, B, and O denote Intro, Verse, Pre-chorus, Chorus, and Outro, respectively. Numerals indicate bar counts.}
\label{tab:app_structures}
\end{table}

\subsection{Harmony Templates}
\label{app:chords}

Chord progressions are selected from musician-curated element-level templates organized by section type, tempo class, and section length. For sections of 16 bars or more, the synthesizer selects an appropriate longer progression or repeats a shorter one. The Chorus progression is also reused in the Intro and Outro. 
The expert synthesizer includes 201 chord-progression templates as shown in Table~\ref{tab:app_chords}. The implementation stores these templates in \texttt{pop\_chordprog.py}.

\begin{table}[htbp]
\centering
\small
\begin{tabular}{llrr}
\toprule
Section & Tempo class & 8-bar & 16-bar \\
\midrule
Verse & Slow & 20 & 9 \\
Verse & Mid & 23 & 4 \\
Verse & Fast & 8 & 9 \\
Pre-chorus & Slow & 13 & -- \\
Pre-chorus & Mid & 23 & -- \\
Pre-chorus & Fast & 14 & -- \\
Chorus & Slow & 9 & 11 \\
Chorus & Mid & 15 & 16 \\
Chorus & Fast & 4 & 23 \\
\midrule
\multicolumn{2}{l}{\textbf{Total}} & \multicolumn{2}{c}{\textbf{201}} \\
\bottomrule
\end{tabular}
\caption{Number of chord progression templates by section type, tempo class, and section length. Dashes (--) indicate that no 16-bar templates are provided for Pre-chorus sections, as they are fixed to 8 bars in our structure design.}
\label{tab:app_chords}
\end{table}

\subsection{Rhythm Generation}
\label{app:rhythm}
Rhythm is generated on an eighth-note grid. Each section contains four phrases: 2-bar phrases for 8-bar sections and 4-bar phrases for 16-bar sections. Phrase syllable counts are sampled from $[5,13]$ for 2-bar phrases and $[8,18]$ for 4-bar phrases, and are then allocated to rhythmic components such as \textit{Pickup}, \textit{Full}, \textit{Unfull}, and \textit{Ending}. Each component type has a distinct rhythmic character: 
\begin{itemize}
\item \textit{Pickup}: may begin with long rests (weak start)
\item \textit{Full}: dense, contains no rests (full bar)
\item \textit{Unfull}: may begin with short rests (light gaps)
\item \textit{Ending}: may end with rests (cadential space)
\end{itemize}

For each component type and syllable count, we pre-enumerate all possible eighth-note patterns that place the specified number of onsets within the component duration while respecting the component's rhythmic character. These patterns are stored in a lookup table. 
During rhythm generation, after fixing a component-type sequence and syllable allocation, the system randomly selects a matching pattern from the corresponding set.

The implementation of these rhythm patterns and allocation rules is available in \texttt{pop\_rhythm.py}.

\subsection{Pitch Generation}
\label{app:pitch}
We generate pitches for each note in two stages: first identifying strong and weak positions, then sampling pitches for strong notes followed by weak notes.

Pitch generation first marks strong positions. A \textit{strong} beat is defined as the downbeat of each bar, plus the last note of every phrase to ensure phrase-final closure. However, a note falling on a strong beat is not always treated as strong: appoggiaturas are considered weak, and the following note takes the strong status.  

The synthesizer then samples strong pitches under register, harmony, and interval-transition constraints.
Remaining weak-note groups are filled using hand-crafted rules conditioned on key/mode, chord, surrounding strong pitches, and group length. 
This two-stage procedure yields melodies that are harmonically coherent and vocally singable. 
The complete implementation, including all transition tables and rule logic, is provided in \texttt{pop\_melody.py}.

\section{Dataset Construction and RWC Preparation}
\label{app:data}

\subsection{Synthetic Corpus}
\label{app:synthetic_corpus}
For the synthetic corpus, the expert synthesizer enumerates twelve major keys, two vocal registers (Male and Female), and three tempo classes (Slow, Mid, and Fast), using a fixed 4/4 meter. For each configuration, a structural template and section-level chord progressions are selected from the corresponding curated pools, then the pipeline constructs paired melody MIDI files and metadata. The same synthesis procedure generates the 10k, 30k, 100k, and 300k datasets used for model-selection experiments.

\subsection{Real-Pop Adaptation Examples}
\label{app:rwc_processing}

Real-pop adaptation uses the RWC Popular Music Database. We exclude two non-4/4 pieces and manually prepare 98 melody-only, half-song MIDI examples covering the intro-to-interlude region. The prepared data are associated with key/mode, tempo class, section, and chord annotations. The preparation normalizes key/mode to C major, records section boundaries by starting bar, and stores beat-aligned chord labels as MIDI markers. Intro and Outro are treated as fixed eight-bar regions without required chord labels in this annotation protocol.

Preprocessing quantizes notes to an eighth-note grid, splits notes crossing bar boundaries, and augments the prepared melodies by transposition across 12 keys, subject to vocal-register constraints. This procedure produces 847 tokenized adaptation examples from the 98 prepared source melodies.

\section{Tokenization, Modeling, and Training}
\label{app:training}

\begin{figure}[htbp]
\centering
\IfFileExists{Figures/fig5_train_pipeline_5.drawio.pdf}{%
  \includegraphics[width=\linewidth]{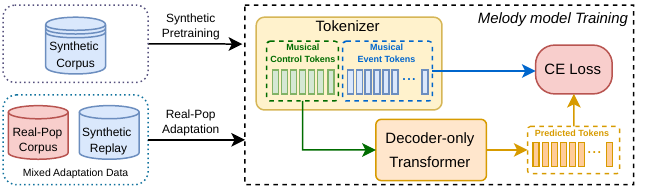}%
}{\PopMissingFigure{Figures/fig5_train_pipeline_5.drawio.pdf}}
\caption{Melody model training process with synthetic pretraining and real-pop adaptation using synthetic replay. }
\label{fig:fig5_train_pipeline}
\end{figure}

As illustrated in Figure~\ref{fig:fig5_train_pipeline}, the training pipeline is separated into event tokenization, synthetic pretraining, and real-pop adaptation with synthetic replay.

\subsection{Event Representation and Conditional Generation}
\label{app:tokens}

\begin{figure}[htbp]
\centering
\IfFileExists{Figures/fig3_tokennizer.drawio_5.pdf}{%
  \includegraphics[width=\linewidth]{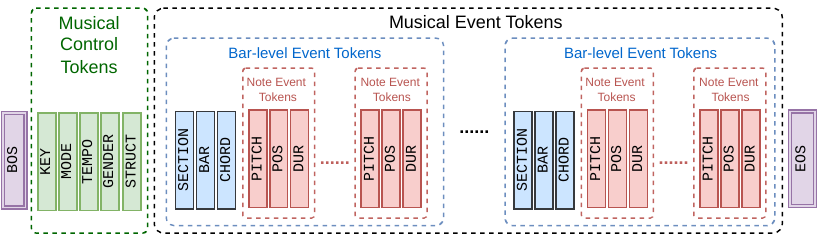}%
}{\PopMissingFigure{Figures/fig3_tokennizer.drawio_5.pdf}}
\caption{Event-based tokenization with a control-token prefix followed by bar-level and note-level events. }
\label{fig:fig3_modeling_training}
\end{figure}

A melody and its metadata are linearized into an event-token sequence, as illustrated in Figure~\ref{fig:fig3_modeling_training}.
The sequence begins with global control tokens for key/mode, tempo class, vocal register, and song structure. 
Each bar then contains section and chord information, followed by note events encoding pitch, onset position, and duration.
This representation enables conditional autoregressive generation with a decoder-only Transformer.

Let $x_i=(x_{i,0},\ldots,x_{i,T_i})$ be a tokenized melody and $x_{i,0}=\mathrm{BOS}$. The model is pretrained with teacher-forced next-token cross-entropy:
\begin{equation}
\mathcal{L}_{\mathrm{pretrain}}(\theta)=-\sum_i\sum_{t=1}^{T_i}\log p_{\theta}(x_{i,t}\mid x_{i,<t}).
\end{equation}
The resulting synthetic-pretrained checkpoint is \textsc{Synth-Base}. Synthetic pretraining covers diverse keys, tempos, structures, vocal registers, and chord progressions, aiming to learn the controllable behaviors of the expert synthesizer.

\subsection{Real-Pop Adaptation with Synthetic Replay}
\label{app:replay}

Synthetic pretraining provides broad musical-condition coverage, but the output distribution of a rule-based system may still differ from naturally composed pop melodies. Directly finetuning a synthetic-pretrained model on a small real-pop corpus may overfit and weaken condition-following behavior. To retain some of the control learned during pretraining, adaptation combines all real-pop examples with a sampled replay subset of the synthetic corpus. Let $\mathcal{D}_{\mathrm{real}}$ and $\mathcal{D}_{\mathrm{syn}}$ be the real and synthetic sets. For replay ratio $\rho$,
\begin{equation}
\begin{split}
\left|\mathcal{D}_{\mathrm{syn}}^{(\rho)}\right| &= \rho\left|\mathcal{D}_{\mathrm{real}}\right|,\\
\mathcal{D}_{\mathrm{adapt}}^{(\rho)} &= \mathcal{D}_{\mathrm{real}}\cup\mathcal{D}_{\mathrm{syn}}^{(\rho)}.
\end{split}
\end{equation}
The final reported \textsc{Real-FT} uses a \emph{real-pop:synthetic replay ratio of 2:1}, corresponding to $\rho=1/2$. The same tokenizer and event vocabulary are used in both training stages.

\subsection{Architecture and Hyperparameters}
\label{app:training_details}

\begin{table}[htbp]
\centering
\small
\begin{tabular}{lcc}
\toprule
Item & Pretraining & Adaptation \\
\midrule
Max context length & 1,300 & 1,300 \\
Hidden size & 512 & 512 \\
Decoder layers & 6 & 6 \\
Attention heads & 8 & 8 \\
Dropout & 0.1 & 0.1 \\
Optimizer & AdamW & AdamW \\
Learning rate & $3\times10^{-4}$ & $3\times10^{-5}$ \\
Batch size & 16 & 16 \\
Epochs & 20 & 20 \\
Train/val/test & 8:1:1 & 8:1:1 \\
Training examples & Synthetic only & Real-pop + replay \\
\bottomrule
\end{tabular}
\caption{Training configuration for both training stages.}
\label{tab:app_train}
\end{table}

\subsection{Inference and Bar-Level Chord Conditioning}
\label{app:chord_inject}

\begin{figure}[htbp]
\centering
\IfFileExists{Figures/fig4_tokennizer.drawio_2.pdf}{%
  \includegraphics[width=\linewidth]{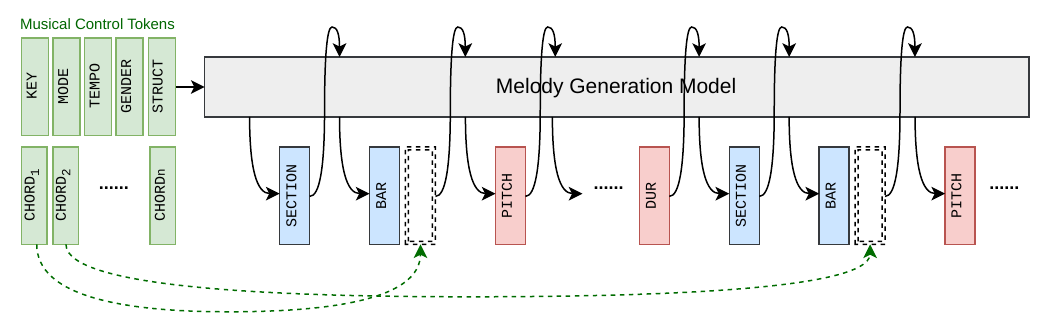}%
}{\PopMissingFigure{Figures/fig4_tokennizer.drawio_2.pdf}}
\caption{Online chord injection: the chord token is inserted at each bar boundary to steer subsequent note-event decoding.}
\label{fig:fig4_chordinjection}
\end{figure}

During neural inference, user-specified global conditions initialize autoregressive decoding.
To enforce the selected chord progression, the corresponding chord token is injected at each bar boundary before generating note events,
as illustrated in Figure~\ref{fig:fig4_chordinjection}.
In contrast, the rule-based Expert backend applies its own harmony and pitch-generation rules.

\section{Evaluation Protocol and Objective Results}
\label{app:objective}

\subsection{Prompt Grid and Metrics}
\label{app:metrics}

The comparison consists of three generation systems: the rule-based \textsc{Expert}, the synthetic-pretrained \textsc{Synth-Base}, and the adapted \textsc{Real-FT}. 

Each model generates 4,320 melodies from an evaluation grid comprising 12 keys, 3 tempo classes, 2 vocal registers, 2 structure families, 6 chord progressions, and 5 samples per condition, for 12,960 outputs across all three systems.

Four musicologically interpretable metrics are used:
\begin{itemize}
\item \textbf{HAR (Harmony Adherence Rate):} harmony alignment, combining chord-tone rates at strong positions and scale-tone rates at weak positions;
\item \textbf{CAD (Cadence Adherence):} whether phrase- or section-final notes satisfy the designated cadential targets;
\item \textbf{RGL (Chorus--Verse Register Lift):} register contrast between Chorus and Verse;
\item \textbf{SMR (Stepwise Motion Rate):} the proportion of local stepwise melodic motion over Verse and Chorus sections.
\end{itemize}
HAR and CAD are control-following metrics for which higher values indicate stronger adherence to the expert-defined constraints. RGL and SMR characterize melodic style and should instead be read relative to the processed real-pop reference. Complete replication requires the implementation-specific thresholds and aggregation procedures contained in the evaluation code.

\subsection{Synthetic Pretraining Scale}
\label{app:scale}

The synthetic scale study compares 10k, 30k, 100k, and 300k examples. Table~\ref{tab:app_scale} reproduces the HAR and CAD values recorded in the scale ablation. Both metrics improve substantially from 10k to 100k, and the 300k setting gives no further HAR gain and lower CAD. Figure 6 visualizes this trend, motivating the selection of \textsc{Synth-100k} as \textsc{Synth-Base}.

\begin{table}[htbp]
\centering
\small
\begin{tabular}{lcc}
\toprule
System & HAR $\uparrow$ & CAD $\uparrow$ \\
\midrule
Expert reference & 1.000 & 1.000 \\
Synth-10k & 0.914 & 0.543 \\
Synth-30k & 0.991 & 0.834 \\
\textbf{Synth-100k} & \textbf{0.998} & \textbf{0.893} \\
Synth-300k & 0.998 & 0.828 \\
\bottomrule
\end{tabular}
\caption{Synthetic pretraining-scale ablation.}
\label{tab:app_scale}
\end{table}

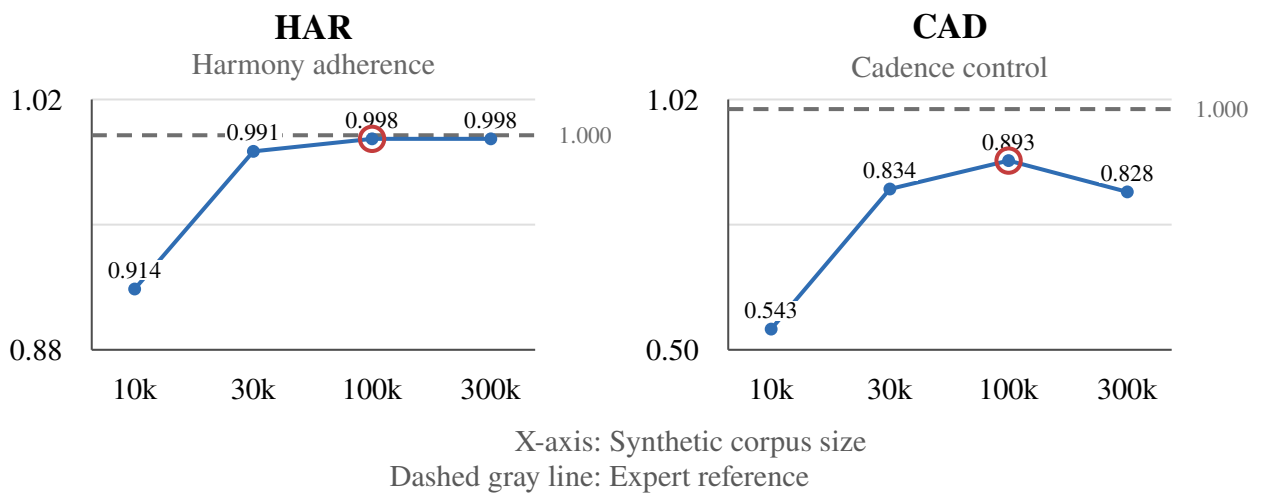
\begin{figure}[htbp]
  \centering
  \definecolor{trendblue}{RGB}{47,109,179}
  \definecolor{realred}{RGB}{196,61,61}
  \definecolor{expertgray}{RGB}{111,111,111}
  \resizebox{\columnwidth}{!}{%
  \begin{tikzpicture}[font=\scriptsize]
    \newcommand{\linepanel}[6]{
      \begin{scope}[shift={(#1,#2)}]
        \pgfmathsetmacro{\ys}{1.75/(#5-#4)}
        \node[font=\scriptsize\bfseries] at (1.55,2.26) {#3};
        \node[font=\tiny,text=black!60] at (1.55,1.98) {#6};
        \draw[gray!25] (0,0) -- (3.10,0);
        \draw[gray!25] (0,0.875) -- (3.10,0.875);
        \draw[gray!25] (0,1.75) -- (3.10,1.75);
        \draw[black!70] (0,0) -- (0,1.75);
        \draw[black!70] (0,0) -- (3.10,0);
        \node[anchor=east,font=\tiny] at (-0.08,0) {#4};
        \node[anchor=east,font=\tiny] at (-0.08,1.75) {#5};
        \foreach \x/\txt in {0.30/10k,1.13/30k,1.96/100k,2.79/300k}
          \node[font=\tiny,anchor=north] at (\x,-0.08) {\txt};
      \end{scope}
    }
    \newcommand{\drawtrend}[8]{
      \draw[trendblue,thick] (0.30,{#1}) -- (1.13,{#2}) -- (1.96,{#3}) -- (2.79,{#4});
      \fill[trendblue] (0.30,{#1}) circle (0.045);
      \fill[trendblue] (1.13,{#2}) circle (0.045);
      \fill[trendblue] (1.96,{#3}) circle (0.045);
      \fill[trendblue] (2.79,{#4}) circle (0.045);
      \node[font=\fontsize{4.7pt}{5.2pt}\selectfont,fill=white,inner sep=0.3pt,anchor=south] at (0.30,{#1+0.055}) {#5};
      \node[font=\fontsize{4.7pt}{5.2pt}\selectfont,fill=white,inner sep=0.3pt,anchor=south] at (1.13,{#2+0.055}) {#6};
      \node[font=\fontsize{4.7pt}{5.2pt}\selectfont,fill=white,inner sep=0.3pt,anchor=south] at (1.96,{#3+0.055}) {#7};
      \node[font=\fontsize{4.7pt}{5.2pt}\selectfont,fill=white,inner sep=0.3pt,anchor=south] at (2.79,{#4+0.055}) {#8};
      \draw[realred,thick] (1.96,{#3}) circle (0.085);
    }
    \newcommand{\drawexpert}[2]{
      \draw[expertgray,densely dashed,thick] (0,{#1}) -- (3.10,{#1});
      \node[anchor=west,font=\fontsize{4.7pt}{5.2pt}\selectfont,text=expertgray] at (3.14,{#1}) {#2};
    }
    \linepanel{0}{0}{HAR}{0.88}{1.02}{Harmony adherence}
    \begin{scope}[shift={(0,0)}]
      \drawexpert{(1.000-0.88)*1.75/(1.02-0.88)}{1.000}
      \drawtrend{(0.914-0.88)*1.75/(1.02-0.88)}{(0.991-0.88)*1.75/(1.02-0.88)}{(0.998-0.88)*1.75/(1.02-0.88)}{(0.998-0.88)*1.75/(1.02-0.88)}{0.914}{0.991}{0.998}{0.998}
    \end{scope}

    \linepanel{4.45}{0}{CAD}{0.50}{1.02}{Cadence control}
    \begin{scope}[shift={(4.45,0)}]
      \drawexpert{(1.000-0.50)*1.75/(1.02-0.50)}{1.000}
      \drawtrend{(0.543-0.50)*1.75/(1.02-0.50)}{(0.834-0.50)*1.75/(1.02-0.50)}{(0.893-0.50)*1.75/(1.02-0.50)}{(0.828-0.50)*1.75/(1.02-0.50)}{0.543}{0.834}{0.893}{0.828}
    \end{scope}
    \node[font=\tiny,text=black!65,align=center] at (3.75,-0.78) {\begin{tabular}{@{}r@{: }l@{}}X-axis & Synthetic corpus size\\Dashed gray line & Expert reference\end{tabular}};
  \end{tikzpicture}}
  \caption{Synthetic pretraining scale ablation on HAR and CAD metrics. The highlighted 100k setting is selected as the \textsc{Synth-Base} checkpoint.}
  \label{fig:synth_scale_ablation}
\end{figure}

\subsection{Replay Mixture Selection}
\label{app:mixture}

Adaptation experiments vary the real-pop:synthetic replay ratio. Table~\ref{tab:app_mix} reproduces values in the mixture ablation. 
Real-only finetuning substantially weakens CAD, whereas adding synthetic replay recovers CAD performance.
Larger replay ratios further strengthen HAR and CAD, but also make the model closer to the synthetic-pretrained behavior. 
Figure 7 shows this trade-off, and we select the 2:1 ratio for the final \textsc{Real-FT} setting because it moves closest to the Real-pop reference while retaining strong harmony and cadence controllability.

\begin{table}[htbp]
\centering
\small
\begin{tabular}{lcc}
\toprule
Real-pop:synthetic replay & HAR $\uparrow$ & CAD $\uparrow$ \\
\midrule
1:0 (real-only) & 0.917 & 0.320 \\
\textbf{2:1 (final Real-FT)} & \textbf{0.983} & \textbf{0.834} \\
1:1 & 0.987 & 0.844 \\
1:2 & 0.994 & 0.890 \\
\midrule
Expert reference & 1.000 & 1.000 \\
Real-pop reference & 0.898 & 0.798 \\
\bottomrule
\end{tabular}
\caption{Real-pop adaptation with synthetic replay. The 2:1 mixture is the reported final checkpoint; the other rows are ablation settings.}
\label{tab:app_mix}
\end{table}

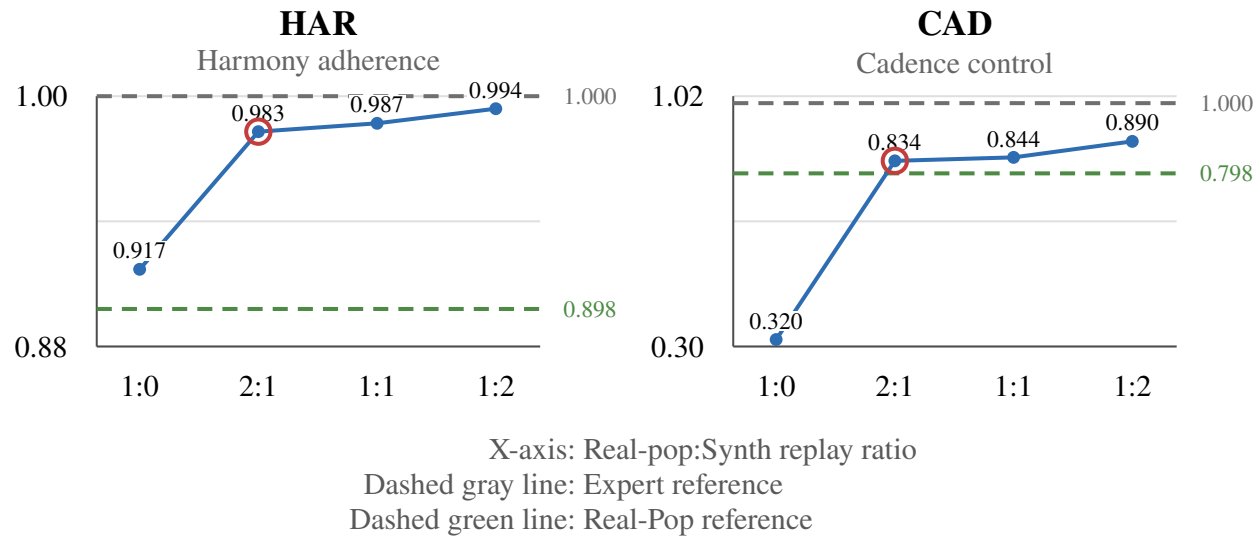
\begin{figure}[htbp]
  \centering
  \definecolor{trendblue}{RGB}{47,109,179}
  \definecolor{refgreen}{RGB}{77,139,69}
  \definecolor{expertgray}{RGB}{111,111,111}
  \definecolor{realred}{RGB}{196,61,61}
  \resizebox{\columnwidth}{!}{%
  \begin{tikzpicture}[font=\scriptsize]
    \newcommand{\linepanel}[6]{
      \begin{scope}[shift={(#1,#2)}]
        \node[font=\scriptsize\bfseries] at (1.55,2.26) {#3};
        \node[font=\tiny,text=black!60] at (1.55,1.98) {#6};
        \draw[gray!25] (0,0) -- (3.10,0);
        \draw[gray!25] (0,0.875) -- (3.10,0.875);
        \draw[gray!25] (0,1.75) -- (3.10,1.75);
        \draw[black!70] (0,0) -- (0,1.75);
        \draw[black!70] (0,0) -- (3.10,0);
        \node[anchor=east,font=\tiny] at (-0.08,0) {#4};
        \node[anchor=east,font=\tiny] at (-0.08,1.75) {#5};
        \foreach \x/\txt in {0.30/1:0,1.13/2:1,1.96/1:1,2.79/1:2}
          \node[font=\tiny,anchor=north] at (\x,-0.08) {\txt};
      \end{scope}
    }
    \newcommand{\drawtrend}[8]{
      \draw[trendblue,thick] (0.30,{#1}) -- (1.13,{#2}) -- (1.96,{#3}) -- (2.79,{#4});
      \fill[trendblue] (0.30,{#1}) circle (0.045);
      \fill[trendblue] (1.13,{#2}) circle (0.045);
      \fill[trendblue] (1.96,{#3}) circle (0.045);
      \fill[trendblue] (2.79,{#4}) circle (0.045);
      \node[font=\fontsize{4.7pt}{5.2pt}\selectfont,fill=white,inner sep=0.3pt,anchor=south] at (0.30,{#1+0.055}) {#5};
      \node[font=\fontsize{4.7pt}{5.2pt}\selectfont,fill=white,inner sep=0.3pt,anchor=south] at (1.13,{#2+0.055}) {#6};
      \node[font=\fontsize{4.7pt}{5.2pt}\selectfont,fill=white,inner sep=0.3pt,anchor=south] at (1.96,{#3+0.055}) {#7};
      \node[font=\fontsize{4.7pt}{5.2pt}\selectfont,fill=white,inner sep=0.3pt,anchor=south] at (2.79,{#4+0.055}) {#8};
      \draw[realred,thick] (1.13,{#2}) circle (0.085);
    }
    \newcommand{\drawref}[2]{
      \draw[refgreen,densely dashed,thick] (0,{#1}) -- (3.10,{#1});
      \node[anchor=west,font=\fontsize{4.7pt}{5.2pt}\selectfont,text=refgreen] at (3.14,{#1}) {#2};
    }
    \newcommand{\drawexpert}[2]{
      \draw[expertgray,densely dashed,thick] (0,{#1}) -- (3.10,{#1});
      \node[anchor=west,font=\fontsize{4.7pt}{5.2pt}\selectfont,text=expertgray] at (3.14,{#1}) {#2};
    }
    \linepanel{0}{0}{HAR}{0.88}{1.00}{Harmony adherence}
    \begin{scope}[shift={(0,0)}]
      \drawexpert{(1.000-0.88)*1.75/(1.00-0.88)}{1.000}
      \drawref{(0.898-0.88)*1.75/(1.00-0.88)}{0.898}
      \drawtrend{(0.917-0.88)*1.75/(1.00-0.88)}{(0.983-0.88)*1.75/(1.00-0.88)}{(0.987-0.88)*1.75/(1.00-0.88)}{(0.994-0.88)*1.75/(1.00-0.88)}{0.917}{0.983}{0.987}{0.994}
    \end{scope}

    \linepanel{4.45}{0}{CAD}{0.30}{1.02}{Cadence control}
    \begin{scope}[shift={(4.45,0)}]
      \drawexpert{(1.000-0.30)*1.75/(1.02-0.30)}{1.000}
      \drawref{(0.798-0.30)*1.75/(1.02-0.30)}{0.798}
      \drawtrend{(0.320-0.30)*1.75/(1.02-0.30)}{(0.834-0.30)*1.75/(1.02-0.30)}{(0.844-0.30)*1.75/(1.02-0.30)}{(0.890-0.30)*1.75/(1.02-0.30)}{0.320}{0.834}{0.844}{0.890}
    \end{scope}
    \node[font=\tiny,text=black!65,align=center] at (3.75,-0.98) {\begin{tabular}{@{}r@{: }l@{}}X-axis & Real-pop:Synth replay ratio\\Dashed gray line & Expert reference\\Dashed green line & Real-Pop reference\end{tabular}};
  \end{tikzpicture}}
  \caption{Real-pop adaptation mixture ablation on HAR and CAD metrics. The highlighted 2:1 mixture is selected as the final \textsc{Real-FT} setting.}
  \label{fig:adapt_mix_ablation}
\end{figure}

\subsection{Comparison of the Three Final Backends}
\label{app:final_metrics}

\begin{figure}[htbp]
  \centering
  \definecolor{expertgray}{RGB}{111,111,111}
  \definecolor{synthblue}{RGB}{47,109,179}
  \definecolor{realred}{RGB}{196,61,61}
  \definecolor{refgreen}{RGB}{77,139,69}
  \resizebox{\columnwidth}{!}{%
  \begin{tikzpicture}[font=\scriptsize]
    \newcommand{\linepanel}[8]{
      \begin{scope}[shift={(#1,#2)}]
        \pgfmathsetmacro{\ys}{1.70/(#5-#4)}
        \node[font=\scriptsize\bfseries] at (1.55,#7) {#3};
        \node[font=\tiny,text=black!60] at (1.55,#8) {#6};
        \draw[gray!25] (0,0) -- (3.10,0);
        \draw[gray!25] (0,0.850) -- (3.10,0.850);
        \draw[gray!25] (0,1.70) -- (3.10,1.70);
        \draw[black!70] (0,0) -- (0,1.70);
        \draw[black!70] (0,0) -- (3.10,0);
        \node[anchor=east,font=\tiny] at (-0.08,0) {#4};
        \node[anchor=east,font=\tiny] at (-0.08,1.70) {#5};
        \foreach \x/\txt in {0.30/Expert,1.13/Synth-base,1.96/Real-FT,2.79/Real-Pop}
          \node[font=\tiny,rotate=25,anchor=east] at ({\x+0.1},-0.08) {\txt};
      \end{scope}
    }
    \newcommand{\drawobjtrend}[8]{
      \draw[expertgray,densely dashed,thick] (0,{#1}) -- (3.10,{#1});
      \draw[refgreen,densely dashed,thick] (0,{#4}) -- (3.10,{#4});
      \draw[black!70,thick] (0.30,{#1}) -- (1.13,{#2}) -- (1.96,{#3}) -- (2.79,{#4});
      \fill[expertgray] (0.30,{#1}) circle (0.045);
      \fill[synthblue] (1.13,{#2}) circle (0.05);
      \fill[realred] (1.96,{#3}) circle (0.05);
      \fill[refgreen] (2.79,{#4}) circle (0.045);
      \node[font=\fontsize{4.7pt}{5.2pt}\selectfont,fill=white,inner sep=0.3pt,anchor=south] at (0.30,{#1+0.055}) {#5};
      \node[font=\fontsize{4.7pt}{5.2pt}\selectfont,fill=white,inner sep=0.3pt,anchor=south] at (1.13,{#2+0.055}) {#6};
      \node[font=\fontsize{4.7pt}{5.2pt}\selectfont,fill=white,inner sep=0.3pt,anchor=south] at (1.96,{#3+0.055}) {#7};
      \node[font=\fontsize{4.7pt}{5.2pt}\selectfont,fill=white,inner sep=0.3pt,anchor=south] at (2.79,{#4+0.055}) {#8};
    }

    \foreach \x/\c/\txt in {0.20/expertgray/Expert (ref.),1.85/synthblue/Synth-Base,3.95/realred/Real-FT,5.75/refgreen/Real-Pop (ref.)} {
      \fill[\c] (\x,-0.88) circle (0.055);
      \node[anchor=west,font=\tiny] at (\x+0.12,-0.88) {\txt};
    }

    \linepanel{0}{3.05}{HAR}{0.88}{1.02}{Harmony adherence}{2.32}{2.02}
    \begin{scope}[shift={(0,3.05)}]
      \drawobjtrend{(1.000-0.88)*1.70/(1.02-0.88)}{(0.998-0.88)*1.70/(1.02-0.88)}{(0.983-0.88)*1.70/(1.02-0.88)}{(0.898-0.88)*1.70/(1.02-0.88)}{1.000}{0.998}{0.983}{0.898}
    \end{scope}

    \linepanel{4.45}{3.05}{CAD}{0.78}{1.02}{Cadence control}{2.32}{2.02}
    \begin{scope}[shift={(4.45,3.05)}]
      \drawobjtrend{(1.000-0.78)*1.70/(1.02-0.78)}{(0.893-0.78)*1.70/(1.02-0.78)}{(0.834-0.78)*1.70/(1.02-0.78)}{(0.798-0.78)*1.70/(1.02-0.78)}{1.000}{0.893}{0.834}{0.798}
    \end{scope}

    \linepanel{0}{0}{RGL}{3.0}{6.3}{Register lift}{2.20}{1.91}
    \begin{scope}[shift={(0,0)}]
      \drawobjtrend{(5.992-3.0)*1.70/(6.3-3.0)}{(5.881-3.0)*1.70/(6.3-3.0)}{(4.946-3.0)*1.70/(6.3-3.0)}{(3.231-3.0)*1.70/(6.3-3.0)}{5.992}{5.881}{4.946}{3.231}
    \end{scope}

    \linepanel{4.45}{0}{SMR}{0.60}{0.75}{Stepwise motion}{2.20}{1.91}
    \begin{scope}[shift={(4.45,0)}]
      \drawobjtrend{(0.634-0.60)*1.70/(0.75-0.60)}{(0.612-0.60)*1.70/(0.75-0.60)}{(0.663-0.60)*1.70/(0.75-0.60)}{(0.724-0.60)*1.70/(0.75-0.60)}{0.634}{0.612}{0.663}{0.724}
    \end{scope}
  \end{tikzpicture}}
  \caption{Objective comparison of \textsc{Expert}, \textsc{Synth-Base}, and \textsc{Real-FT} across four symbolic metrics.
Dashed lines indicate the \textsc{Expert} and \textsc{Real-Pop} reference values.}
  \label{fig:obj_metrics}
\end{figure}
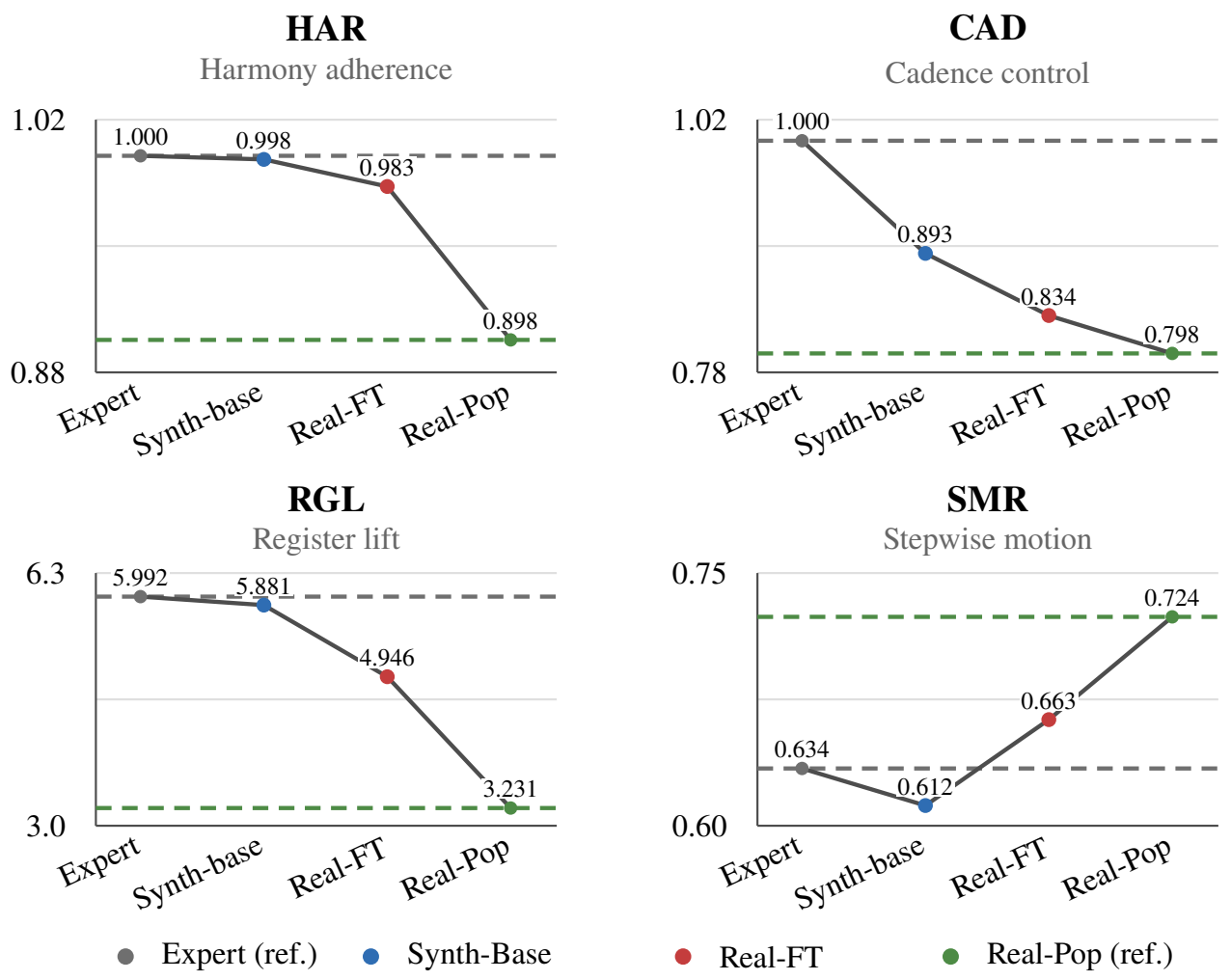

The final-checkpoint values of objective evaluation are shown in Table~\ref{tab:app_finalobj} and Figure~\ref{fig:obj_metrics}. 
Two expected behaviors are demonstrated.
First, \textsc{Synth-Base} closely matches the \textsc{Expert} reference on all four metrics, indicating that the Transformer successfully learns the musical characteristics encoded by the expert synthesizer through large-scale synthetic pretraining.
Second, after real-pop adaptation, \textsc{Real-FT} shifts consistently toward \textsc{Real-Pop} on all four metrics, suggesting that finetuning effectively adapts the synthetic-pretrained model toward real-pop melodic characteristics while preserving much of the controllability acquired during synthetic pretraining.

\begin{table}[htbp]
\centering
\small
\begin{tabular}{lcccc}
\toprule
System & HAR $\uparrow$ & CAD $\uparrow$ & RGL & SMR \\
\midrule
Expert & 1.000 & 1.000 & 5.992 & 0.634 \\
Synth-Base (100k) & 0.998 & 0.893 & 5.881 & 0.612 \\
Real-FT (2:1) & 0.983 & 0.834 & 4.946 & 0.663 \\
\midrule
RWC real-pop reference & 0.898 & 0.798 & 3.231 & 0.724 \\
\bottomrule
\end{tabular}
\caption{Objective evaluation for the three final systems and the descriptive real-pop reference. RGL and SMR are not monotonic quality scores.}
\label{tab:app_finalobj}
\end{table}

\section{Human Evaluation}
\label{app:human}

For human evaluation, 30 participants evaluated 18 generated songs from 6 prompt groups and 3 systems using a 5-point MOS scale and a best-of-three preference task. As shown in Table~\ref{tab:app_human}, \textsc{Real-FT} achieved the highest MOS (3.548) and preference rate (48.3\%).

\begin{table}[htbp]
\centering
\small
\begin{tabular}{lcc}
\toprule
System & MOS $\uparrow$ & Best-of-three $\uparrow$ \\
\midrule
Expert & $3.231\pm0.059$ & 25.6\% \\
Synth-Base & $3.244\pm0.064$ & 26.1\% \\
\textbf{Real-FT} & $\mathbf{3.548\pm0.057}$ & \textbf{48.3\%} \\
\bottomrule
\end{tabular}
\caption{Human study with 30 participants and 18 generated songs. MOS is mean$\pm$SE on a 5-point scale; Best-of-three is the preference win rate within each prompt group.}
\label{tab:app_human}
\end{table}

\section{Similarity Check Details}
\label{app:similarity_check}

\begin{figure}[htbp]
\centering
\IfFileExists{Figures/similarity_violin_Expert.csv}{%
  \IfFileExists{Figures/similarity_violin_SynthBase.csv}{%
    \IfFileExists{Figures/similarity_violin_RealFT.csv}{%

\centering
\definecolor{simgray}{RGB}{111,111,111}
\definecolor{simblue}{RGB}{47,109,179}
\definecolor{simred}{RGB}{196,61,61}
\resizebox{\columnwidth}{!}{%
\begin{tikzpicture}
\begin{axis}[
  width=6.0cm,
  height=3.2cm,
  xmin=0.45,
  xmax=3.55,
  ymin=0,
  ymax=0.6,
  ylabel={Nearest-neighbor similarity},
  xtick={1,2,3},
  xticklabels={\textsc{Expert},\textsc{Synth-Base},\textsc{Real-FT}},
  xticklabel style={font=\scriptsize},
  yticklabel style={font=\scriptsize},
  label style={font=\scriptsize},
  title={Similarity to Real-Pop Reference},
  title style={font=\scriptsize\bfseries},
  ymajorgrids=true,
  xmajorgrids=false,
  grid style={gray!20},
  axis on top,
  clip=false,
]
\addplot[draw=simgray, fill=simgray, fill opacity=0.28, no marks] table[x=x, y=y, col sep=comma] {Figures/similarity_violin_Expert.csv};
\addplot[draw=simblue, fill=simblue, fill opacity=0.28, no marks] table[x=x, y=y, col sep=comma] {Figures/similarity_violin_SynthBase.csv};
\addplot[draw=simred, fill=simred, fill opacity=0.28, no marks] table[x=x, y=y, col sep=comma] {Figures/similarity_violin_RealFT.csv};
\end{axis}
\end{tikzpicture}}

    }{\PopMissingFigure{Figures/similarity_violin_RealFT.csv}}%
  }{\PopMissingFigure{Figures/similarity_violin_SynthBase.csv}}%
}{\PopMissingFigure{Figures/similarity_violin_Expert.csv}}
\caption{Nearest-neighbor similarity between melodies generated by the three systems and the RWC reference set.}
\label{fig:app_similarity_distribution}
\end{figure}
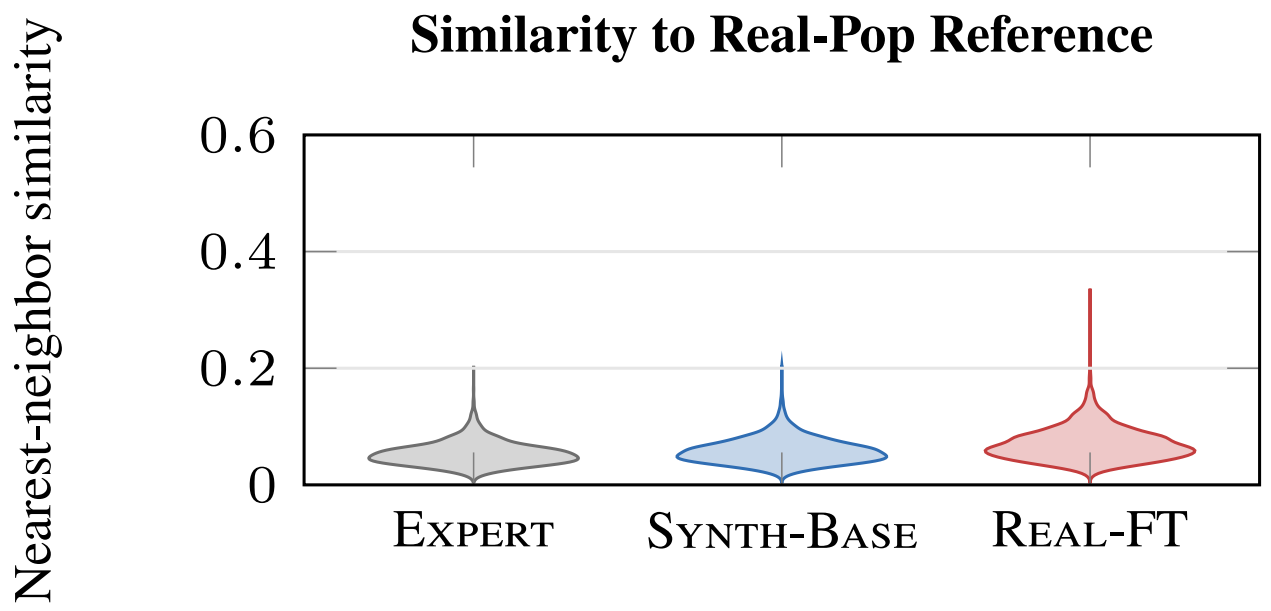

For each generated melody, we extract transposition-invariant interval $n$-grams and retrieve its nearest match from the selected RWC reference set. 
Figure~\ref{fig:app_similarity_distribution} shows the nearest-neighbor similarity distributions for 12,960
generated melodies across the three systems.
Manual inspection of the highest-scoring pairs revealed no obvious near-duplicate melodies. 

Although this analysis is based on a limited real-pop reference set, it provides empirical evidence for
assessing melodic similarity and offers a useful reference for evaluating potential near-copy risks in real-world applications.

\section{Demonstration and Limitations}
\label{app:demo}

\subsection{Creator and Researcher Workflows}

\begin{figure}[htbp]
\centering
\IfFileExists{Figures/UI_aaai_1.png}{%
  \includegraphics[width=0.99\linewidth]{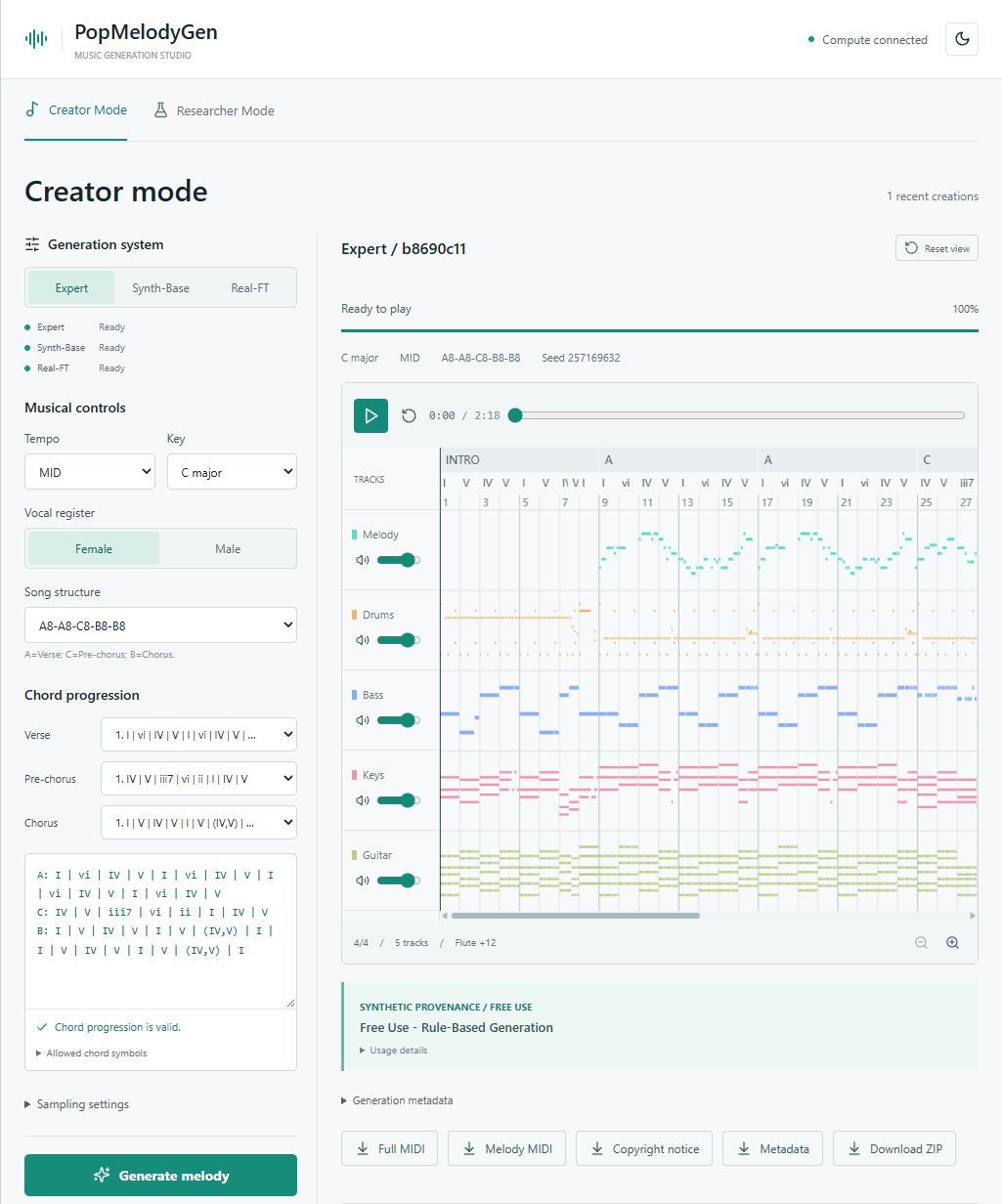}%
}{\PopMissingFigure{Figures/UI_aaai_1.png}}
\caption{Creator Mode interface for controllable melody generation, audition, and export.}
\label{fig:fig5_UI_1}
\end{figure}

\begin{figure}[htbp]
\centering
\IfFileExists{Figures/UI_aaai_2.png}{%
  \includegraphics[width=0.99\linewidth]{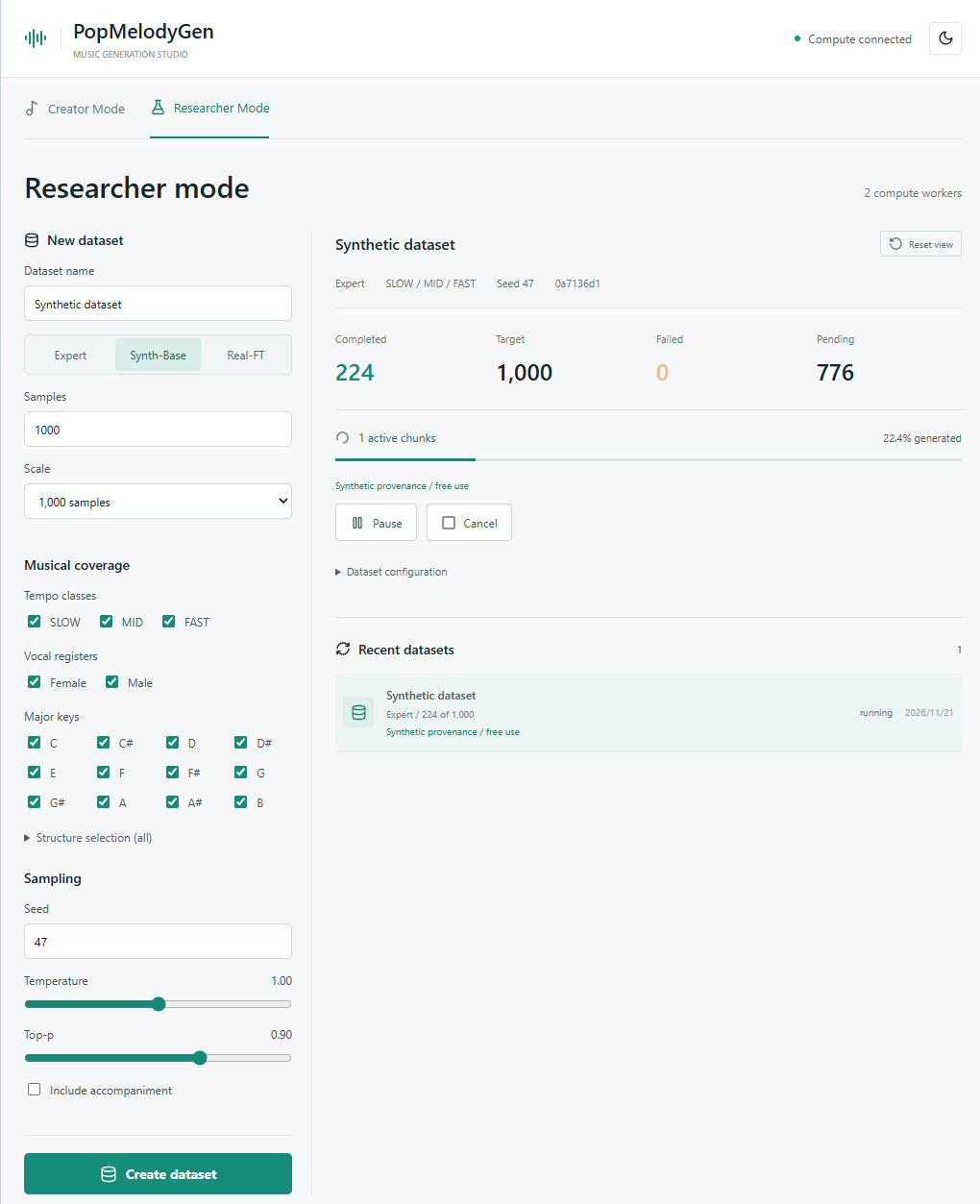}%
}{\PopMissingFigure{Figures/UI_aaai_2.png}}
\caption{Researcher Mode interface for batch generation of annotated symbolic melodies.}
\label{fig:fig5_UI_2}
\end{figure}

Figures~\ref{fig:fig5_UI_1} and~\ref{fig:fig5_UI_2} illustrate the two workflows of the deployed platform.
Creator Mode supports interactive melody generation, audition, and export, while Researcher Mode enables
batch generation of labeled MIDI--metadata pairs.
The Node.js frontend communicates with the generation backends through an inference API.

\subsection{Scope and Limitations}
The expert synthesizer relies on predefined musical structures, curated chord templates, and simplified
phrase-level rules. 
While these constraints facilitate controllability and reproducibility, they limit stylistic diversity. 
Moreover, real-pop adaptation uses a relatively small corpus, restricting its coverage of musical styles
and structures. Finally, the neural model focuses on symbolic melody generation rather than full-song
production. Although the platform provides accompaniment for audition, lyrics, singing voices, and learned
multi-track arrangement remain outside the current modeling scope.

\end{document}